\documentclass[aps,prl,reprint,superscriptaddress,longbibliography,floatfix]{revtex4-2}
\usepackage[T1]{fontenc}
\usepackage{amsmath,amssymb,bm,graphicx,booktabs,microtype,xcolor}
\usepackage[colorlinks=true,allcolors=blue]{hyperref}
\usepackage{orcidlink}
\hypersetup{pdftitle={Resolving the spinon hierarchy of the Heisenberg antiferromagnet},pdfauthor={Ian P. McCulloch; Jesse J. Osborne}}
\graphicspath{{figures/}}
\newcommand{\dd}{\mathrm{d}}
\newcommand{\ii}{\mathrm{i}}
\DeclareMathOperator{\sech}{sech}

\begin{document}
\title{Resolving the spinon hierarchy of the Heisenberg antiferromagnet}
\author{Ian P. McCulloch\,\orcidlink{0000-0002-8983-6327}}
\email[Contact author: ]{ian@phys.nthu.edu.tw}
\affiliation{Department of Physics, National Tsing Hua University, Hsinchu 30013, Taiwan}
\author{Jesse J. Osborne\,\orcidlink{0000-0003-0415-0690}}
\affiliation{Max Planck Institute of Quantum Optics, 85748 Garching, Germany}
\affiliation{Department of Physics and Arnold Sommerfeld Center for Theoretical
Physics (ASC), Ludwig Maximilian University of Munich, 80333 Munich, Germany}
\affiliation{Munich Center for Quantum Science and Technology (MCQST), 80799
Munich, Germany}
\date{30 October 2026}
\begin{abstract}
A local spin fluctuation in the spin-$\tfrac12$ Heisenberg antiferromagnet
creates an even number of fractional spinon excitations. We resolve its
zero-temperature dynamical structure factor through six spinons directly
in the thermodynamic limit. Beyond the established two-spinon contribution
of $72.89\%$, the four-spinon
response gives approximately $21.90\%$ of the integrated intensity and
six spinons contribute a further $4.3\%$, bringing the response to about
$99\%$ of both the intensity and first-frequency-moment sum rules.
We derive momentum-independent first-moment fractions in the calculated
sectors and a general upper-edge law
$S_n\sim A_n(k)(\omega_{n,u}-\omega)^{(n^2-3)/2}$ for every even $n$ at
nonzero momentum. The rapid growth of the edge exponent explains why
the expanding multiparticle bandwidth carries very little high-energy
weight, while higher sectors remain appreciable within the lower-sector
continua.
\end{abstract}
\maketitle

The spin-$\tfrac12$ Heisenberg antiferromagnetic chain is a paradigmatic
example of fractionalization: a local spin flip creates spin-$\tfrac12$
spinons rather than a single magnon \cite{faddeev1981}. Its excitation
threshold \cite{descloizeaux1962} and multiparticle continua make exact
many-body physics directly accessible to spectroscopy. Neutron
experiments have measured absolutely normalized spectra and established
weight beyond the two-spinon contribution \cite{mourigal2013,lake2013}.
Higher-spinon states contribute substantial weight within the two-spinon
band and extend the response to higher energies. The quantitative question
is how a local probe distributes its intensity among successive spinon sectors.

Two complementary exact approaches address this question.
Vertex operators supply infinite-chain form factors
\cite{jimbo1992,jimbomiwa,bougourzi1997,abada1997}; the exact two-spinon
response accounts for $72.89\%$ of the integrated longitudinal intensity
\cite{bougourzi1996,karbach1997}. Caux and Hagemans evaluated the
four-spinon continuum and reported near-saturation by two and four
spinons \cite{caux2006}. Correcting the four-spinon normalization and numerical integration,
we find that the two- and four-spinon sectors together account for approximately $94.8\%$
of the intensity, leaving appreciable weight in higher sectors.
Finite-chain determinant form factors
\cite{kitanine1999} instead enable spectral sums
\cite{cauxmaillet2005,caux2005}, developed into the ABACUS framework
\cite{caux2009}. Their thermodynamic limits provide a further connection
between these approaches \cite{kitanine2019,kitanine2021}.
Time-dependent density-matrix renormalization and matrix-product-state
(MPS) methods offer an independent route to the full response
\cite{white2004,daley2004,haegeman2011}. Absolute normalization, rather than a fitted
overall amplitude, is essential for comparing these descriptions.

Here we resolve the thermodynamic response through six spinons. We
derive the four-spinon normalization independently, evaluate its spectrum
on a million-point grid, and calculate six-spinon spectra at eight
momenta. Two and four spinons leave about $5\%$ of the weight, reduced to
about $1\%$ after six spinons. A general upper-edge law explains why increasing
spinon number expands the bandwidth much faster than the visible
high-energy response.
Finite-chain sector weights exhibit an approximately logarithmic drift,
with no clear onset of saturation even at $N=1024$. At $k=\pi/2$, two and
four spinons still exhaust $99.73\%$ of the finite-chain first moment. Nevertheless,
infinite-MPS time evolution resolves the thermodynamic six-spinon contribution,
despite its finite bond-dimension correlation length.

\begin{figure*}[!t]
 \includegraphics[width=\textwidth]{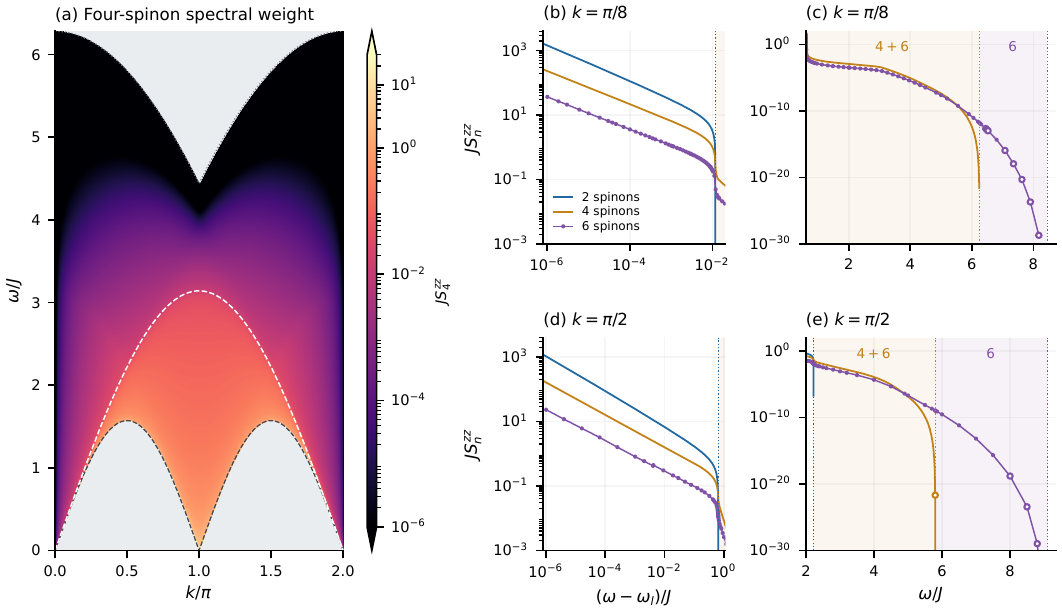}
 \caption{Unbroadened thermodynamic spinon spectra, with no fitted normalization.
 (a) Four-spinon heat map from $1{,}048{,}576$ directly evaluated
 points: 1024 momenta in $0<k<\pi$, each with 512 frequencies below
 and 512 above the two-spinon upper boundary (white dashed line).
 The other half-zone is reflected.
 (b,c) Sector contributions at $k=\pi/8$; (d,e) at $k=\pi/2$.
 Left panels use logarithmic distance from $\omega_l=(\pi/2)\sin k$;
 right panels use linear frequency to resolve the upper bands.
 The windows overlap, with logarithmic intensity and different vertical ranges.
 Colored dotted lines mark sector upper boundaries; shading identifies the
 contributing sectors among $S_2,S_4,S_6$.
 Six-spinon calculations use rapidity cutoff $R=1040$. Open symbols
 mark points that miss the four-spinon quadrature-order change target of $0.1\%$,
 or the six-spinon sampling criteria including a $5\%$ relative-error target.}
 \label{fig:spectra}
\end{figure*}

\textit{Conventions and calculation.---}
We set $J$, the lattice spacing, and $\hbar$ to one, with
$H=\sum_j\bm S_j\cdot\bm S_{j+1}$. The longitudinal structure factor is
\begin{equation}
 \begin{aligned}
 S^{zz}(k,\omega)&=\sum_j e^{-\ii k j}\int_{-\infty}^{\infty}\dd t\,
 e^{\ii\omega t}\langle S_j^z(t)S_0^z(0)\rangle\\
 &=\sum_{n=2,4,\ldots}S_n(k,\omega).
 \end{aligned}
 \label{eq:dsf}
\end{equation}
All sector densities have this same absolute normalization. Define
\begin{equation}
 \begin{aligned}
 m_{r,n}(k)&=\int_0^\infty\frac{\dd\omega}{2\pi}\,\omega^r S_n(k,\omega),\\
 I_{r,n}&=\int_0^{2\pi}\frac{\dd k}{2\pi}\,m_{r,n}(k).
 \end{aligned}
 \label{eq:moments}
\end{equation}
The exact sum rules \cite{hohenberg1974,karbach1997} are $I_0=1/4$ and
$m_1(k)=\tfrac23(\ln2-\tfrac14)(1-\cos k)$, hence
$I_1=\tfrac23(\ln2-\tfrac14)$.

\begin{samepage}
A spinon has $p\in(0,\pi)$, energy $e(p)=\pi\sin p/2$, and rapidity
$\beta=\ln\tan(p/2)$. With $E=\sum_j e(p_j)$ and $P=\sum_jp_j$, write
\begin{equation}
 S_n(k,\omega)=\int_{(0,\pi)^n}\dd^np\,W_n(\bm p)
 \delta_{2\pi}(k-P)\delta(\omega-E).
 \label{eq:integral}
\end{equation}
\end{samepage}
Here $W_n$ includes the longitudinal normalization and a single $1/n!$
for labelled momenta; $\delta_{2\pi}$ retains every momentum branch.
We evaluate the four-spinon phase-space integrals by deterministic
quadrature. For six spinons, we use randomized quasi-Monte Carlo
integration, with coordinates adapted to soft pairs and clusters.
Independent randomizations of the sampling sequences provide statistical
error estimates.

The six-spinon form factors involve two auxiliary contour integrals
and fifteen polarization components. We evaluate the contour integrals
as residue sums with controlled tails, cross-checking independent
implementations. Global moments are integrated independently of the
frequency slices. The companion article gives the methods, convergence
studies and complete normalization derivation \cite{companion}.

The End Matter summarizes our independent derivation of the four-spinon
normalization from finite-anisotropy amplitudes, yielding half the
coefficient printed in Ref.~\cite{caux2006}. This factor-of-two correction
was previously identified by Suzuki \cite{suzuki2026private}.
An Ising-limit check and finite-chain matrix elements independently
support the result without fitting a sum rule.

Separately, an audit of the thermodynamic four-spinon routines
distributed with ABACUS identifies numerical integration losses that
partially masked the normalization error. Restoring the missed weight
while retaining the printed coefficient would make the combined
two- and four-spinon contributions exceed the exact sum rules.
The companion article gives the numerical reconciliation and the
scope of this source-code comparison \cite{companion}.

\textit{Spectra and sum-rule accounting.---}
Figure~\ref{fig:spectra}(a) resolves the four-spinon continuum
across the Brillouin zone, including weight above the two-spinon band
and rapid suppression toward its own upper edge.
We calculated six-spinon slices at $k=j\pi/8$, $j=1,\ldots,8$;
the full spectra and convergence studies are reported in
the companion article \cite{companion}.

At $k=\pi/8$ [Fig.~\ref{fig:spectra}(b,c)], the two-spinon continuum
occupies a narrow interval of width $0.0118J$, while the four- and
six-spinon contributions extend over a much broader energy range.
The higher-spinon spectra also exhibit a shoulder near
$\omega/J=\pi\cos(k/2)\simeq3.0812$. This feature marks an internal
boundary of the integration domain \cite{caux2006}, where soft spinons
carry net momentum $\pi$ and the remaining hard pair reaches its
upper energy boundary. At $k=\pi/2$, this feature coincides with the
two-spinon upper edge.

At $k=\pi/2$ [Fig.~\ref{fig:spectra}(d,e)],
four and six spinons both carry substantial weight inside the two-spinon
continuum. Above its upper edge, the two higher-sector curves retain
similar shapes over a broad interval: between $\omega=2.3$ and $4$,
$S_4$ falls by a factor of $184$, while $S_6/S_4$ rises only from
$0.298$ to $0.383$. Six spinons overtake four near $\omega\simeq5$,
where both contributions are already faint. The four-spinon boundary is
$\omega_{4,u}=2\pi\cos(\pi/8)\simeq5.80491$, whereas
$\omega_{6,u}=3\pi\cos(\pi/12)\simeq9.10364$.
The response within that additional six-spinon bandwidth is extremely
small. The other slices show the same broad shape resemblance, with
momentum- and frequency-dependent precision \cite{companion}.

Table~\ref{tab:weights} shows the details of the global spectral weights.
The direct six-spinon integral uses the dimensionless rapidity cutoff
$\max_j|\beta_j|\le R=1040$. Since a spinon has energy
$e(\beta)=\pi J/(2\cosh\beta)$, large $|\beta|$ describes very soft
spinons with momenta approaching $0$ or $\pi$. The cutoff therefore
excludes configurations containing extremely soft spinons, which can
accompany finite-energy spinons throughout the spectrum.
The six-spinon contributions are
$4.2809(108)\%$ of the intensity and $4.6456(98)\%$ of the first moment, where
the parentheses denote observed sampling errors in the final digits.
% Alternative momentum quadratures are consistent with these estimates.
% The rapidity cutoff omits very soft configurations, not merely the
% high-frequency tail seen in Fig.~\ref{fig:spectra}(e).
Extrapolating the soft-tail beyond the cutoff predicts leading additions of $0.05457$
and $0.05601$ percentage points to the intensity and first moment,
respectively. Including that
unfitted leading tail gives our best full-domain estimates of $99.12\%$
and $99.05\%$ for two plus four plus six spinons.
% Uniform exhaustion of
% the leading soft regions is not yet established, so these estimates do
% not have a rigorous omitted-tail bound. The finite-cutoff values and
% sampling errors remain separately specified.
% The unfilled fraction
% of the direct calculation includes both the omitted six-spinon tail
% and sectors with eight or more spinons.

\begin{table}[!tbp]
\caption{Percentages of the exact global sum rules. Parentheses on
finite-cutoff rows give six-spinon sampling errors only. The final row
adds the independently predicted leading rapidity tail; its model
uncertainty is distinct from sampling error.}
\label{tab:weights}
\begin{ruledtabular}
\begin{tabular}{lrr}
Contribution & Intensity & First moment\\
Two & 72.8913 & 71.3030\\
Four & 21.8981 & 23.0421\\
Six, $R=1040$ & 4.2809(108) & 4.6456(98)\\
Two + four + six, $R=1040$ & 99.0703(108) & 98.9908(98)\\
With leading-tail estimate & 99.12 & 99.05\\
\end{tabular}
\end{ruledtabular}
\end{table}

\begin{figure}[!tbp]
 \includegraphics[width=\columnwidth]{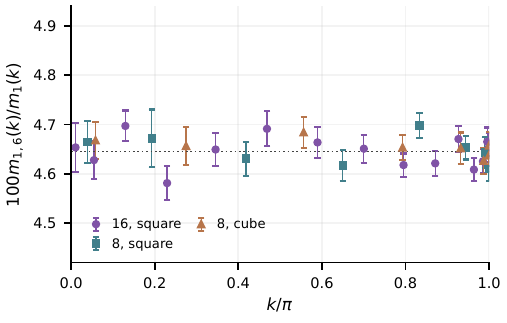}
 \caption{Six-spinon contribution to the first-moment sum rule as a function
of momentum. Symbols distinguish three momentum quadrature rules;
the legend gives their node counts and endpoint mappings.
Error bars show sampling uncertainty. The dotted line is the global
first-moment fraction obtained from the 16-node quadrature.
All results use rapidity cutoff $R=1040$.}
 \label{fig:moments}
\end{figure}

\textit{Momentum dependence and soft spinons.---}
The two-, four- and six-spinon sectors each carry a fixed fraction
of the first-moment sum rule, independent of momentum:
\begin{equation}
m_{1,n}(k)=I_{1,n}(1-\cos k),\qquad n=2,4,6.
\label{eq:boostmoment}
\end{equation}
This identity follows by integrating a common shift of all rapidities:
the energy weighting cancels the momentum Jacobian, and the momentum
branches contribute equally. The derivation in the End Matter assumes
the full rapidity domain. A finite cutoff can introduce deviations
from this dependence, which Figure~\ref{fig:moments} tests numerically.

Soft-spinon limits also connect the spectral shapes of successive sectors.
In the six-spinon sector, a pair becoming soft with total momentum zero modulo
$2\pi$ leaves a four-spinon response at leading order.
A four-soft-spinon
cluster similarly leaves a two-spinon core \cite{companion}.
These limits offer a mechanism for the resemblance between the
four- and six-spinon spectra and explain the slowly decaying rapidity
tails. The resemblance is not uniform across the spectrum:
$S_6/S_4$ varies strongly near the four-spinon upper edge.

\begin{figure}[!tbp]
 \includegraphics[width=\columnwidth]{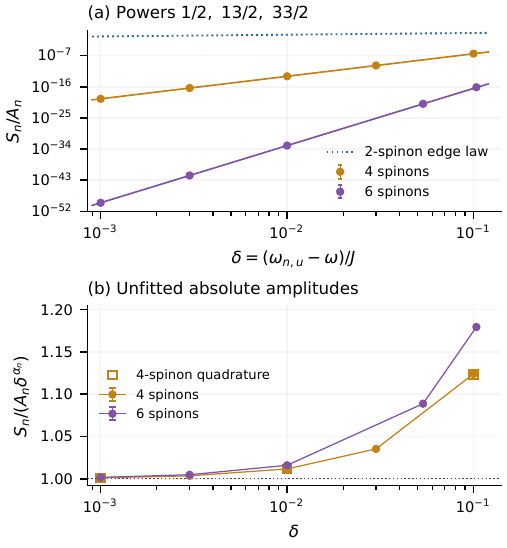}
 \caption{Upper-edge hierarchy at $k=\pi/2$. (a) Direct four- and
 six-spinon integrations divided by their analytically determined
 coefficients, compared with the powers $13/2$ and $33/2$; the dotted
 two-spinon curve shows its established leading $1/2$ law.
 (b) The same data divided by the complete unfitted leading asymptotes.
 Open squares are independent four-spinon quadrature results.
 Monte Carlo sampling errors are smaller than the symbol size.
 These compact energy-shell calculations have no soft-rapidity
 cutoff and are separate from the full-band data in Fig.~\ref{fig:spectra}.}
 \label{fig:edges}
\end{figure}

\textit{A general upper-edge hierarchy.---}
For fixed $0<k\le\pi$, the upper boundary of an even-$n$ sector is
\begin{equation}
 \omega_{n,u}(k)=\frac{n\pi}{2}
 \max_{s:\,0<k+2\pi s<n\pi}\sin\!\left(\frac{k+2\pi s}{n}\right).
 \label{eq:edge}
\end{equation}
The maximum within each momentum branch occurs at equal momenta.
Writing $p_i=p_*+x_i$, with $\sum_i x_i=0$, makes the energy deficit
quadratic in the remaining $n-1$ coordinates. The fixed-momentum energy
shell therefore contributes a factor
$\delta^{(n-3)/2}$, where $\delta=\omega_{n,u}-\omega$.
This density of states is weighted by the squared spin form factor,
which vanishes as the spinon momenta coalesce. Each pair-transition
factor supplies a quadratic zero. After these factors are removed,
the residual polarization vector is regular and nonzero at the
equal-momentum maximum, giving
\begin{equation}
W_n\sim B_n(k)\prod_{i<j}(x_i-x_j)^2,\qquad B_n(k)>0.
\label{eq:vandermonde}
\end{equation}
The End Matter establishes the required regularity and nonvanishing
for all even $n$. Since each momentum difference scales as
$\sqrt{\delta}$, the $n(n-1)/2$ pair factors contribute
$\delta^{n(n-1)/2}$. Combining this with the phase-space factor
$\delta^{(n-3)/2}$ yields
\begin{equation}
 S_n(k,\omega)\sim A_n(k)\delta^{\alpha_n},\qquad
 \alpha_n=\frac{n^2-3}{2},\quad A_n(k)>0.
 \label{eq:power}
\end{equation}

Figure~\ref{fig:edges} tests both powers and coefficients using the full
form factors on exact energy shells. At distance $\delta = 10^{-3}$, the ratios
to the leading predictions are $1.0011463$ for four spinons (deterministic
quadrature) and $1.0015795(59)$ for six spinons. The six-spinon amplitude
$A_6(\pi/2)\simeq4.81\times10^{-21}$ in our $J=1$ convention is
very small, showing that the outer tail is suppressed by both a small amplitude and
the power $33/2$. The general result predicts $61/2$ and $97/2$ for
eight and ten spinons without requiring their full spectral integrals.

\begin{figure}[!tbp]
 \includegraphics[width=\columnwidth]{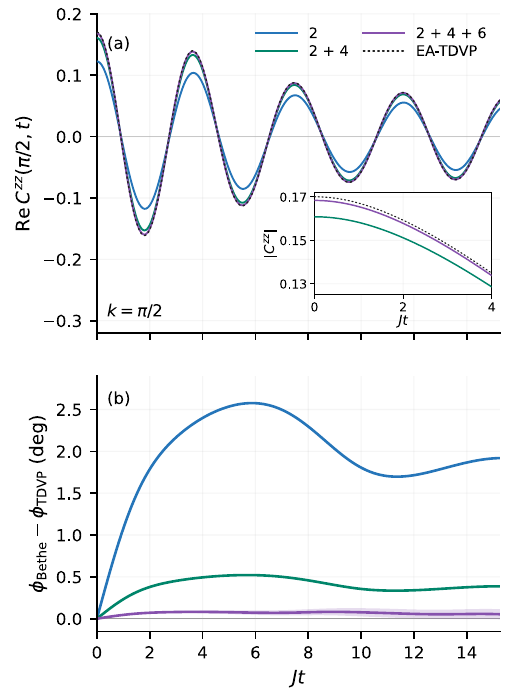}
 \caption{Independent time-domain comparison at $k=\pi/2$.
 (a) Real part of $C^{zz}(k,t)$ from momentum-space EA-TDVP (black
 dotted) and the absolutely normalized partial spinon sums (solid).
 The inset shows the complex magnitude for $0\le Jt\le4$;
 the two-spinon curve lies below its displayed range.
 (b) Phase differences relative to TDVP, unchanged by positive amplitude
 rescaling. Shading denotes six-spinon sampling errors. No
 amplitude, phase or energy shift is fitted.}
 \label{fig:validation}
\end{figure}

\textit{Finite-size flow and independent checks.---}
Finite-chain calculations provide an independent check of the
four-spinon normalization. Using determinant form factors
\cite{kitanine1999,caux2005} with string deviations retained
\cite{hagemans2007}, we compare selected four-spinon matrix elements
with matched two-spinon matrix elements. Their ratios approach the
thermodynamic predictions with our normalization to within $0.4\%$
at the largest sizes.

The integrated sector weights converge much more slowly.
At $N=1024$ and $k=\pi/2$, our four-spinon catalogue contains
$8{,}355{,}776$ states. Together, two and four spinons account for
$99.73\%$ of the exact finite-chain first moment. Their separate
contributions, however, are $89.73\%$ and $10.00\%$, far from the
thermodynamic values of $71.30\%$ and $23.04\%$
\cite{companion}. Near-saturation of a finite-chain sum rule can
therefore conceal substantial redistribution between spinon sectors
and gives no comparable bound on higher-sector weight in the
infinite chain.

Over the accessible sizes, the sector fractions drift approximately
linearly with $\log N$, without a clear onset of saturation.
Configurations involving increasingly soft spinons redistribute
weight as the system grows. A controlled asymptotic description
of this redistribution remains open, even though individual
finite-energy matrix elements already reproduce the thermodynamic
normalization.

\textit{Resolving higher-spinon dynamics with MPS.---}
Momentum-space MPS time evolution provides an independent probe of
the higher-spinon response. At $k=\pi/2$, we use excitation-ansatz
time evolution based on an adaptation of TDVP
\cite{haegeman2016,vandamme2022}, starting from a U(1)-symmetric
ground-state MPS of bond dimension 200 (End Matter).
We compare the resulting signal with the spectral transforms
$C_{\le n}(k,t)=\sum_{m=2,4,\ldots,n}\int_0^\infty
\dd\omega\,S_m(k,\omega)e^{-\ii\omega t}/(2\pi)$.
Adding four and then six spinons successively improves agreement
with the TDVP signal [Fig.~\ref{fig:validation}(a)].
The amplitude inset resolves the separation between the sector sums.

The phase provides a stringent test of the spectral distribution,
since an overall intensity rescaling cannot change it.
Writing $\Delta\phi_{\le n}=\arg[C_{\le n}C_{\rm TDVP}^*]$,
the maximum phase discrepancy over $0\le Jt\le15.25$ decreases
from $2.58^\circ$ for two spinons to $0.52^\circ$ for two plus four,
and to $0.082^\circ$ after including six
[Fig.~\ref{fig:validation}(b)].
An independent frequency-grid transform confirms agreement at
approximately the $0.1^\circ$ level (End Matter).
The six-spinon contribution therefore improves the time dependence
as well as the absolute intensity.

At zero time, six spinons contribute about $20\%$ as much as four,
while the remaining difference from TDVP is about $22\%$ of the
six-spinon contribution. This is consistent in scale with a
continuing hierarchy of higher-spinon weights.
Further numerical
convergence tests are needed to separate that contribution from
the remaining integration and MPS errors.

\textit{Discussion and outlook.---}
Resolving the response through six spinons accounts for approximately
$99\%$ of both the intensity and first-moment sum rules, while revealing
substantial higher-spinon weight within the familiar two-spinon band.
Positivity of the omitted spectral weight bounds its contribution
to real-space correlations at every separation and time by the
missing global intensity, thus keeping the absolute error from
truncating the spinon expansion small even at long times.

The general upper-edge hierarchy explains why the increasing
multiparticle bandwidth produces little visible high-energy intensity:
the six-spinon outer edge already vanishes as the $33/2$ power.
Higher-spinon contributions are therefore most readily detected through
absolute intensities and detailed line shapes within the bright
continuum.
The absolutely normalized spectra provide quantitative predictions
for spectroscopy of Heisenberg-chain materials
\cite{mourigal2013,lake2013}. Our MPS comparison also demonstrates
a sensitive signature of six spinons in the time-dependent amplitude
and phase. This suggests a complementary experimental route:
time-resolved measurements could distinguish higher-spinon dynamics
from an overall rescaling of the dominant two-spinon response.

The finite-size results reveal a different obstacle to resolving
fractionalization. A nearly saturated finite-chain sum rule can coexist
with sector weights far from their thermodynamic values. Their
approximately logarithmic drift persists to chains over a
thousand sites, making the distribution among spinon sectors much
harder to extrapolate than the total response. Understanding how
increasingly soft spinons produce this redistribution is an open
analytical problem.

The spectral representation also opens a route to dynamics beyond the
times accessible to direct MPS evolution. Our TDVP calculation reaches
$Jt\simeq 15$, while the spectral approach allows sector-resolved
spin correlations and linear responses to wave packets with chosen
spatial or momentum envelopes to be evaluated at much later times.
This provides a resource for investigating the propagation, spreading and
interference of fractionalized excitations. Resolving these dynamics
by spinon number makes it possible to identify how successive
particle sectors contribute, including at times where direct
time evolution becomes computationally demanding.
This complements MPS approaches based
on infinite boundary conditions and dynamical windows
\cite{phien2012,phien2013,milsted2013,zauner2015}, and studies of
quasiparticle wave-packet propagation and scattering
\cite{vandamme2021}.

\begin{acknowledgments}
We thank Junji Suzuki and Jean-S\'ebastien Caux for helpful discussions
and correspondence concerning the four-spinon normalization and calculation.
The authors acknowledge the use of OpenAI's GPT-6-Astra for assistance
throughout this work. Responsibility for the scientific analysis, verification,
and conclusions rests with the authors.
I.P.M. acknowledges funding from the Taiwan National Science and Technology Council (NSTC) Grant
No.~115-2112-M-007-050-MY3.
J.J.O. acknowledges funding by the Max Planck Society, the Deutsche
Forschungsgemeinschaft (DFG, German Research Foundation) under Germany’s
Excellence Strategy -- EXC-2111 -- 390814868, and the European Research Council
(ERC) under the European Union’s Horizon Europe research and innovation program
(Grant Agreement No.~101165667)--ERC Starting Grant QuSiGauge.
The Bethe ansatz calculations were performed using the DiCOS distributed cloud
resources of the Academia Sinica Grid Computing Centre (ASGC), supported by Academia Sinica, Taiwan.
\end{acknowledgments}

\paragraph*{Data and code.}
The calculation uses the Bethe library \cite{bethecode}.
The four-spinon heat map and six-spinon momentum slices will be
made freely available through a Zenodo data release
\cite{spinondata}.

\bibliography{references}

@book{jimbomiwa,
 author={Jimbo, Michio and Miwa, Tetsuji},
 title={Algebraic Analysis of Solvable Lattice Models},
 series={CBMS Regional Conference Series in Mathematics}, volume={85},
 publisher={American Mathematical Society}, year={1994}, doi={10.1090/cbms/085}}

@article{abada1997,
 author={Abada, A. and Bougourzi, A. H. and Si-Lakhal, B.},
 title={Exact four-spinon dynamical correlation function of the {Heisenberg} model},
 journal={Nuclear Physics B}, volume={497}, pages={733--753}, year={1997},
 doi={10.1016/S0550-3213(97)00285-X}, url={https://arxiv.org/abs/hep-th/9702028}}

@article{karbach1997,
 author={Karbach, Michael and M{\"u}ller, Gerhard and Bougourzi, A. H. and Fledderjohann, Andreas and M{\"u}tter, K.-H.},
 title={Two-spinon dynamic structure factor of the one-dimensional {$s=1/2$ Heisenberg} antiferromagnet},
 journal={Physical Review B}, volume={55}, pages={12510--12517}, year={1997},
 doi={10.1103/PhysRevB.55.12510}}

@article{caux2006,
 author={Caux, Jean-S{\'e}bastien and Hagemans, Rob},
 title={The four-spinon dynamical structure factor of the {Heisenberg} chain},
 journal={Journal of Statistical Mechanics: Theory and Experiment},
 volume={2006}, pages={P12013}, year={2006}, doi={10.1088/1742-5468/2006/12/P12013},
 url={https://arxiv.org/abs/cond-mat/0611319}}

@article{bougourzi1997,
 author={Bougourzi, A. H.}, title={Exact {$n$}-spinon dynamic structure factor of the {Heisenberg} model},
 journal={Modern Physics Letters B}, volume={10}, pages={1237--1244}, year={1996},
 doi={10.1142/S0217984996001401}, url={https://arxiv.org/abs/q-alg/9701003}}

@incollection{tarasov1998,
 author={Tarasov, Vitaly}, title={Completeness of the Hypergeometric Solutions of the {qKZ} Equation at Level Zero},
 booktitle={L. D. Faddeev's Seminar on Mathematical Physics},
 series={American Mathematical Society Translations, Series 2}, volume={201},
 publisher={American Mathematical Society}, pages={309--321}, year={2000},
 url={https://arxiv.org/abs/math/9811048}, note={Theorem 4.4 in the arXiv version}}

@misc{bethecode,
  author = {{Uni20 Bethe project}},
  title = {Bethe-ansatz calculations: implementation and reproduction material},
  year = {2026},
  url = {https://github.com/Uni20-dev/bethe}
}

@article{daley2004,
   author={Daley, A. J. and Kollath, C. and Schollw{\"o}ck, U. and Vidal, G.},
   title={Time-dependent density-matrix renormalization-group using adaptive effective {Hilbert} spaces},
   journal={Journal of Statistical Mechanics: Theory and Experiment},
   volume={2004},
   pages={P04005},
   year={2004},
   doi={10.1088/1742-5468/2004/04/P04005},
   url={https://arxiv.org/abs/cond-mat/0403313}
}

@misc{suzuki2026private,
   author={Suzuki, Junji},
   howpublished={private communication}, year={2026}
}

@article{phien2012,
   author = {Phien, Ho N. and Vidal, Guifr{\'e} and McCulloch, Ian P.},
   title = {Infinite boundary conditions for matrix product state calculations},
   journal = {Physical Review B},
   volume = {86},
   pages = {245107},
   year = {2012},
   doi = {10.1103/PhysRevB.86.245107}
}

@article{phien2013,
   author = {Phien, Ho N. and Vidal, Guifr{\'e} and McCulloch, Ian P.},
   title = {Dynamical windows for real-time evolution with matrix product states},
   journal = {Physical Review B},
   volume = {88},
   pages = {035103},
   year = {2013},
   doi = {10.1103/PhysRevB.88.035103}
}

@article{milsted2013,
   author = {Milsted, Ashley and Haegeman, Jutho and Osborne, Tobias J.
           and Verstraete, Frank},
   title = {Variational matrix product ansatz for nonuniform dynamics
          in the thermodynamic limit},
   journal = {Physical Review B},
   volume = {88},
   pages = {155116},
   year = {2013},
   doi = {10.1103/PhysRevB.88.155116}
}

@article{zauner2015,
   author = {Zauner, Valentin and Ganahl, Martin and Evertz, Hans Gerd
           and Nishino, Tomotoshi},
   title = {Time evolution within a comoving window: scaling of signal
          fronts and magnetization plateaus after a local quench
          in quantum spin chains},
   journal = {Journal of Physics: Condensed Matter},
   volume = {27},
   pages = {425602},
   year = {2015},
   doi = {10.1088/0953-8984/27/42/425602}
}

@misc{companion,
   author={McCulloch, Ian P.},
   title={Four- and six-spinon dynamics of the Heisenberg antiferromagnet},
   year={2026},
   note={Companion article in preparation}
}

@misc{spinondata,
 title = {Numerical data for {Resolving the spinon hierarchy
          of the Heisenberg antiferromagnet}},
 year = {2026},
 howpublished = {Zenodo},
 note = {Data release in preparation; DOI to be assigned}
}

@misc{caux2014,
  author = {Caux, Jean-S{\'e}bastien},
  title = {Comment on `{Revisiting the Exact Dynamical Structure Factor of the Heisenberg Antiferromagnetic Model}' by {A. H. Bougourzi}},
  year = {2014},
  howpublished = {arXiv:1402.4793},
  url = {https://arxiv.org/abs/1402.4793}
}

@article{faddeev1981,
 author={Faddeev, L. D. and Takhtajan, L. A.},
 title={What is the spin of a spin wave?},
 journal={Physics Letters A}, volume={85}, pages={375--377}, year={1981},
 doi={10.1016/0375-9601(81)90335-2}}

@article{descloizeaux1962,
 author={des Cloizeaux, Jacques and Pearson, J. J.},
 title={Spin-Wave Spectrum of the Antiferromagnetic Linear Chain},
 journal={Physical Review}, volume={128}, pages={2131--2135}, year={1962},
 doi={10.1103/PhysRev.128.2131}}

@article{mourigal2013,
 author={Mourigal, M. and Enderle, M. and Kl{\"o}pperpieper, A. and Caux, J.-S. and Stunault, A. and R{\o}nnow, H. M.},
 title={Fractional spinon excitations in the quantum {Heisenberg} antiferromagnetic chain},
 journal={Nature Physics}, volume={9}, pages={435--441}, year={2013},
 doi={10.1038/nphys2652}, url={https://arxiv.org/abs/1306.4678}}

@article{lake2013,

 author={Lake, B. and Tennant, D. A. and Caux, J.-S. and Barthel, T. and Schollw{\"o}ck, U. and Nagler, S. E. and Frost, C. D.},
 title={Multispinon Continua at Zero and Finite Temperature in a Near-Ideal {Heisenberg} Chain},
 journal={Physical Review Letters}, volume={111}, pages={137205}, year={2013},
 doi={10.1103/PhysRevLett.111.137205}, url={https://arxiv.org/abs/1307.4071}}

@article{jimbo1992,
 author={Jimbo, Michio and Miki, Kei and Miwa, Tetsuji and Nakayashiki, Atsushi},
 title={Correlation functions of the {XXZ} model for {$\Delta<-1$}},
 journal={Physics Letters A}, volume={168}, pages={256--263}, year={1992},
 doi={10.1016/0375-9601(92)91128-E}, url={https://arxiv.org/abs/hep-th/9205055}}

@article{bougourzi1996,
 author={Bougourzi, A. H. and Couture, M. and Kacir, M.},
 title={Exact two-spinon dynamical correlation function of the one-dimensional {Heisenberg} model},
 journal={Physical Review B}, volume={54}, pages={R12669--R12672}, year={1996},
 doi={10.1103/PhysRevB.54.R12669}, url={https://arxiv.org/abs/q-alg/9604019}}

@article{kitanine1999,
 author={Kitanine, N. and Maillet, J. M. and Terras, V.},
 title={Form factors of the {XXZ Heisenberg spin-1/2} finite chain},
 journal={Nuclear Physics B}, volume={554}, pages={647--678}, year={1999},
 doi={10.1016/S0550-3213(99)00295-3}, url={https://arxiv.org/abs/math-ph/9807020}}

@article{cauxmaillet2005,
 author={Caux, Jean-S{\'e}bastien and Maillet, Jean Michel},
 title={Computation of Dynamical Correlation Functions of {Heisenberg} Chains in a Magnetic Field},
 journal={Physical Review Letters}, volume={95}, pages={077201}, year={2005},
 doi={10.1103/PhysRevLett.95.077201}, url={https://arxiv.org/abs/cond-mat/0502365}}

@article{caux2005,
 author={Caux, Jean-S{\'e}bastien and Hagemans, Rob and Maillet, J. M.},
 title={Computation of dynamical correlation functions of {Heisenberg} chains: the gapless anisotropic regime},
 journal={Journal of Statistical Mechanics: Theory and Experiment},
 volume={2005}, pages={P09003}, year={2005},
 doi={10.1088/1742-5468/2005/09/P09003}, url={https://arxiv.org/abs/cond-mat/0506698}}

@article{caux2009,
 author={Caux, Jean-S{\'e}bastien},
 title={Correlation functions of integrable models: A description of the {ABACUS} algorithm},
 journal={Journal of Mathematical Physics}, volume={50}, pages={095214}, year={2009},
 doi={10.1063/1.3216474}, url={https://arxiv.org/abs/0908.1660}}

@article{hagemans2007,
 author={Hagemans, Rob and Caux, Jean-S{\'e}bastien},
 title={Deformed strings in the {Heisenberg} model},
 journal={Journal of Physics A: Mathematical and Theoretical},
 volume={40}, pages={14605--14647}, year={2007},
 doi={10.1088/1751-8113/40/49/001}, url={https://arxiv.org/abs/0707.2803}}

@article{kitanine2019,
 author={Kitanine, Nikolai and Kulkarni, Giridhar},
 title={Thermodynamic limit of the two-spinon form factors for the zero field {XXX} chain},
 journal={SciPost Physics}, volume={6}, pages={076}, year={2019},
 doi={10.21468/SciPostPhys.6.6.076}, url={https://arxiv.org/abs/1903.09058}}

@article{kitanine2021,
 author={Kitanine, Nikolai and Kulkarni, Giridhar},
 title={Form Factors of the {Heisenberg} Spin Chain in the Thermodynamic Limit: Dealing with Complex {Bethe} Roots},
 journal={SIGMA}, volume={17}, pages={112}, year={2021},
 doi={10.3842/SIGMA.2021.112}, url={https://arxiv.org/abs/2105.10244}}

@article{white2004,
 author={White, Steven R. and Feiguin, Adrian E.},
 title={Real-Time Evolution Using the Density Matrix Renormalization Group},
 journal={Physical Review Letters}, volume={93}, pages={076401}, year={2004},
 doi={10.1103/PhysRevLett.93.076401}, url={https://arxiv.org/abs/cond-mat/0403310}}

@article{haegeman2011,
 author={Haegeman, Jutho and Cirac, J. Ignacio and Osborne, Tobias J. and Pi{\v z}orn, Iztok and Verschelde, Henri and Verstraete, Frank},
 title={Time-Dependent Variational Principle for Quantum Lattices},
 journal={Physical Review Letters}, volume={107}, pages={070601}, year={2011},
 doi={10.1103/PhysRevLett.107.070601}, url={https://arxiv.org/abs/1103.0936}}

@article{hohenberg1974,
 author={Hohenberg, P. C. and Brinkman, W. F.},
 title={Sum rules for the frequency spectrum of linear magnetic chains},
 journal={Physical Review B}, volume={10}, pages={128--131}, year={1974},
 doi={10.1103/PhysRevB.10.128}}

@article{nakayashiki1999,
 author={Nakayashiki, A. and Pakuliak, S. and Tarasov, V.},
 title={On solutions of the {KZ} and {qKZ} equations at level zero},
 journal={Annales de l'Institut Henri Poincar{\'e}, Physique th{\'e}orique},
 volume={71}, pages={459--496}, year={1999},
 url={https://numdam.org/item/AIHPA_1999__71_4_459_0/}}

@article{haegeman2016,
 author={Haegeman, Jutho and Lubich, Christian and Oseledets, Ivan and Vandereycken, Bart and Verstraete, Frank},
 title={Unifying time evolution and optimization with matrix product states},
 journal={Physical Review B}, volume={94}, pages={165116}, year={2016},
 doi={10.1103/PhysRevB.94.165116}}

@article{vandamme2021,
 author={{Van Damme}, Maarten and Vanderstraeten, Laurens and {De Nardis}, Jacopo and Haegeman, Jutho and Verstraete, Frank},
 title={Real-time scattering of interacting quasiparticles in quantum spin chains},
 journal={Physical Review Research}, volume={3}, pages={013078}, year={2021},
 doi={10.1103/PhysRevResearch.3.013078}, url={https://arxiv.org/abs/1907.02474}}

@article{vandamme2022,
 author={{Van Damme}, Maarten and Vanderstraeten, Laurens},
 title={Momentum-resolved time evolution with matrix product states},
 journal={Physical Review B}, volume={105}, pages={205130}, year={2022},
 doi={10.1103/PhysRevB.105.205130}}

@article{haegeman2012,
 author={Haegeman, Jutho and Pirvu, Bogdan and Weir, David J. and Cirac, J. Ignacio and Osborne, Tobias J. and Verschelde, Henri and Verstraete, Frank},
 title={Variational matrix product ansatz for dispersion relations},
 journal={Physical Review B}, volume={85}, pages={100408(R)}, year={2012},
 doi={10.1103/PhysRevB.85.100408}}

@article{haegeman2013,
 author={Haegeman, Jutho and Osborne, Tobias J. and Verstraete, Frank},
 title={Post-matrix product state methods: To tangent space and beyond},
 journal={Physical Review B}, volume={88}, pages={075133}, year={2013},
 doi={10.1103/PhysRevB.88.075133}}

@article{osborne2025,
 author={Osborne, Jesse J. and McCulloch, Ian P.},
 title={Efficient and systematic calculation of arbitrary observables for the matrix product state excitation ansatz},
 journal={Physical Review Research}, volume={7}, pages={023018}, year={2025},
 doi={10.1103/PhysRevResearch.7.023018}}

@misc{mcculloch2024,
 author={McCulloch, Ian P. and Osborne, Jesse J.},
 title={Comment on ``Controlled Bond Expansion for Density Matrix Renormalization Group Ground State Search at Single-Site Costs'' ({Extended Version})},
 year={2024}, eprint={2403.00562v2}, archivePrefix={arXiv}, primaryClass={cond-mat.str-el},
 url={https://arxiv.org/abs/2403.00562v2}}
\clearpage
% Compact technical support for the Letter. Full derivations are developed
% in the companion article; source links are recorded in evidence.md.
\raggedbottom
\section*{End Matter}
\label{em:start}
\setcounter{equation}{0}
\renewcommand{\theequation}{E\arabic{equation}}
\renewcommand{\theHequation}{endmatter.\arabic{equation}}

\textit{Absolute normalization.---}
Let $X_{n,i}$ be the finite-anisotropy local $S^+$ vertex amplitude
\cite{jimbo1992,jimbomiwa,abada1997}, with vacuum label $i=0,1$ and
$q=-e^{-\epsilon}$. Completeness integrates the squared spectral
parameter $\xi_j^2$ once around its circle \cite{caux2014}.
Writing $\xi_j=\ii e^{-\ii\phi_j}$, the integrated transverse weight is
\begin{equation}
 \mathcal W^+_{n,\epsilon}=\frac{1}{2n!}\sum_{i,\mathrm{spin}}
 \int_{[-\pi/2,\pi/2)^n}\prod_j\frac{\dd\phi_j}{\pi}
 |X_{n,i}|^2.
 \label{em:measure}
\end{equation}
The vacuum average and label permutations are separate factors;
retaining both translation channels does not double this measure.
For $\phi_j=\epsilon\beta_j/\pi$,
$\dd\phi_j/\pi=(\epsilon/\pi^2)\cosh\beta_j\,\dd p_j$.
Define the vacuum-independent limit
$D_n=\lim_{\epsilon\downarrow0}\epsilon^n
\prod_j\cosh\beta_j\sum_{\mathrm{spin}}|X_{n,i}|^2$.
The vertex Hamiltonian is $2UH_\Delta U^\dagger$, where
\begin{equation*}
 H_\Delta=\sum_j\bigl(S_j^xS_{j+1}^x+S_j^yS_{j+1}^y
                    +\Delta S_j^zS_{j+1}^z\bigr),
 \qquad \Delta=\cosh\epsilon,
\end{equation*}
and $U$ rotates every other site by $\pi$ about the $z$ axis.
Time rescaling and
$S^{+-}=2S^{zz}$ at the isotropic endpoint give the physical density
in Eq.~\eqref{eq:integral}:
\begin{equation}
 W_n=\frac{(2\pi)^2}{2n!\pi^{2n}}D_n,\qquad n=2,4.
 \label{em:ledger}
\end{equation}
Here the remaining $1/2$ is the energy-delta Jacobian; the operator
$\sigma^+=(\sigma^x+\ii\sigma^y)/2=S^+$ needs no extra spin factor.

Put $B=\Gamma(1/4)|A(\ii\pi/2)|$, $\rho_j=\beta_j/(2\pi)$,
and $\rho_{12}=\rho_1-\rho_2$.
The functions $A$, $I$, $g_l$ and $\mathcal J$ use exactly the
transition-function and four-spinon kernel conventions of
Ref.~\cite{caux2006}; no kernel or domain rescaling is made.
At fixed distinct rapidities, dividing by two two-spinon amplitudes
cancels the per-particle products. With
$X_{2,i}^{ab}=X_{2,i}(\beta_a,\beta_b)$, the residue reduction gives
\begin{equation}
 \frac{|X_{4,i,l}|^2}{|X_{2,i}^{12}X_{2,i}^{34}|^2}
 \longrightarrow\frac{\pi^2|g_l|^2}{16B^8}
 \prod_{\substack{a=1,2\\b=3,4}}e^{-I(\rho_b-\rho_a)},
 \label{em:relative}
\end{equation}
where $l$ is the positive spin label in the four-spinon amplitude.
The two-spinon limit and Eq.~\eqref{em:relative} then yield
\begin{equation}
 D_2=\frac{\pi^2}{4}e^{-I(\rho_{12})},\qquad
 D_4=\frac{\pi^6}{256B^8}\mathcal J.
 \label{em:rawlimits}
\end{equation}
Equation~\eqref{em:ledger} yields $W_2=e^{-I}/4$, reproducing the
known two-spinon response \cite{bougourzi1996,karbach1997} after the two momentum preimages are summed,
and $W_4=\mathcal J/(3072B^8)$, half of that used in Ref.~\cite{caux2006}.
The contour limit and a separate uniform-integrability argument for
the global zeroth and first moments are given in Ref.~\cite{companion}.
Independent spin-basis perturbation theory at the Ising end gives
$\mathcal W_4^+=g^4/16+O(g^6)$, $g=\Delta^{-1}$, agreeing with
Eq.~\eqref{em:measure}. Selected finite-chain matrix-element ratios
provide the isotropic check discussed in the main text. Neither check
sets a fitted normalization.

\textit{First moment and momentum branches.---}
Anchor one labelled rapidity, $b=\beta_n$, and write
$\beta_j=b+\delta_j$, $\delta_n=0$. This transformation has unit
Jacobian and introduces no additional permutation factor. The contour
identities for $n=2,4,6$ give a nonnegative, shift-invariant kernel:
\begin{equation}
 W_n(\bm p(\bm\beta))\prod_j\sech\beta_j
 =\Phi_n(\bm\delta)\sin^2(P/2).
 \label{eq:boost}
\end{equation}
For fixed differences, $P(b)$ increases from zero to $n\pi$, with
$\dd P/\dd b=2E(b)/\pi>0$. At $0<k<2\pi$ the periodic delta function
therefore has $n/2$ roots, $P(b_s)=k+2\pi s$,
$s=0,\ldots,n/2-1$. Setting $E_s=E(b_s)$ gives
\begin{equation}
 m_{j,n}(k)=\frac{\sin^2(k/2)}4
 \int_{\mathbb R^{n-1}}\dd^{n-1}\delta\,\Phi_n(\bm\delta)
 \sum_{s=0}^{n/2-1}E_s^{j-1}.
 \label{em:branches}
\end{equation}
The factor $1/4=(2\pi)^{-1}(\pi/2)$ combines the frequency measure
with the momentum Jacobian. For $j=1$ every branch contributes one:
$m_{1,n}=nZ_n\sin^2(k/2)/8$, where $Z_n=\int\Phi_n\,\dd^{n-1}\delta$.
Its momentum average is $I_{1,n}=nZ_n/16$, proving
Eq.~\eqref{eq:boostmoment}. The zeroth-moment fraction need not be
constant. At finite cutoff only roots in
$-R-\min_j\delta_j\le b_s\le R-\max_j\delta_j$ survive;
their number depends on $k$. The unrestricted identity is consequently
not imposed on the finite-cutoff data in Fig.~\ref{fig:moments}.

\textit{Nonzero coalescence coefficient.---}
The precise statement needed for Eq.~\eqref{eq:power} is that the
residual polarization vector, after extracting the pair-transition
factors, is analytic and nonzero at $\beta_1=\cdots=\beta_n=b$ when
$np=k+2\pi s$ and $k\ne0$ modulo $2\pi$. Positivity alone is insufficient.
We prove nonvanishing by showing that the cycle representing the
physical amplitude lies outside the kernel of the hypergeometric
integral map. The contour identification is developed fully in
Ref.~\cite{companion}.

Put $n=2m$, $r=m-1$, $Z=e^b$ and $Q=Z^2$.
Gamma reflection, reversal of the labels and the auxiliary shift
$t=\alpha+\ii\pi/2$ identify the physical contour map with the
level-zero hypergeometric map \cite{nakayashiki1999,tarasov1998}, with
$z_j=\beta_{n+1-j}$, $h=\ii\pi$, step $2h$ and $\mu=0$. Its condition
$z_i-z_j+h\notin2h\mathbb Z$ permits exact coincidence.
The separating contours remain regular there. Clearing
$\prod_a(y_a-1)^n$, $y_a=e^{t_a-b}$, represents the physical cycle,
up to a nonzero scalar, by
\begin{equation}
 U(Q)=\bigwedge_{j=0}^{m-2}(e_j+Qe_{j+1}),\qquad
 e_j(y)=y^{2j+1}.
 \label{em:cycle}
\end{equation}
For $n=2$ the exterior product is empty. The numerator space consists
of polynomials of degree below $n$.
Tarasov's exact kernel theorem, Theorem 4.4, says the null cycles are
generated by wedging with a one-form and a two-form. At coincidence
the one-form is $v=2\sum_{j=0}^{m-1}\binom n{2j+1}e_j$.
Writing $(y+1)^n=\mathcal E(y)+\mathcal O(y)$ for its even and odd
parts, the cleared two-form is
\begin{equation}
 N_2(x,y)=4\frac{x\mathcal O(x)\mathcal E(y)
                    -y\mathcal E(x)\mathcal O(y)}{x+y}.
 \label{em:secondkernel}
\end{equation}
The apparent pole is removable, and its odd total degree gives one
even and one odd factor in every term.

\begin{samepage}
Let $\pi_O$ project onto
all-odd forms, and define the coefficient functional
\begin{equation}
 \mathcal L(\eta)=[e_0\wedge\cdots\wedge e_{m-1}]
                  (v\wedge\pi_O\eta).
 \label{em:functional}
\end{equation}
\end{samepage}
It annihilates the entire kernel: the first image gives $v\wedge v=0$,
and the second has no all-odd component. On Eq.~\eqref{em:cycle},
$\mathcal L(U)=2(-1)^r d_n(Q)$, where
\begin{equation}
 \begin{aligned}
 d_n(Q)&=\sum_{j=0}^{m-1}(-Q)^j\binom n{2j+1},\\
 d_n(Z^2)&=\frac{(1+Z^2)^m}{Z}\sin(np/2),\qquad p=2\arctan Z.
 \end{aligned}
 \label{em:witness}
\end{equation}
At the physical maximum $\sin(np/2)=(-1)^s\sin(k/2)\ne0$.
The cycle is outside the kernel, proving the required nonzero vector.
Regularity also controls angular directions with partial collisions.

The pair factor is $e^{-I(\rho)}=K\rho^2[1+o(1)]$, $K>0$,
and $\rho'(p_*)=(2\pi\sin p_*)^{-1}$ is finite. Writing
$p_i=p_*+x_i$, $\sum_i x_i=0$, gives
$\delta=(\pi\sin p_*/4)\sum_i x_i^2+O(|\bm x|^3)$.
Rescaling $x_i=\sqrt\delta\,y_i$ therefore yields
\begin{equation}
 S_n\sim A_n(k)\,
 \delta^{\,n(n-1)/2+(n-3)/2},\qquad A_n(k)>0.
 \label{em:edgecount}
\end{equation}
The squared-Vandermonde angular integral is finite and positive;
degenerate maxima contribute additively. This is a fixed-nonzero-$k$
limit and does not exchange the edge limit with $k\to0$.

\textit{Numerical accuracy and the independent comparison.---}
We calculate global moments directly at $R=260,520,1040$,
independently of the eight frequency slices. A 16-node momentum
quadrature is compared with two eight-node alternatives.
Sampling errors are estimated from independent randomizations,
retaining correlations between frequencies, moments and cutoffs
when combining results.

We check the form factors by increasing the arithmetic precision
and refining the residue sums. Frequency-grid refinement and
alternative momentum quadratures test the numerical integration.
The rapidity-cutoff dependence is compared with the analytical
leading tail described below. These checks are assessed separately
from the sampling errors shown in the figures.

The known soft regions predict an omitted contribution $T_j/R$,
with coefficients fixed by lower-sector moments:
\begin{equation}
 \begin{aligned}
 T_0&=8I_{1,2}I_{0,4}+4I_{1,4}I_{0,2},\\
 T_1&=12I_{1,2}I_{1,4}.
 \end{aligned}
 \label{em:tail}
\end{equation}
These yield the separate leading-tail estimates in Table~\ref{tab:weights}.
The measured $520<R\le1040$ shell divided by
$T_j(1/520-1/1040)$ is $1.01224(594)$ for intensity and
$1.00921(473)$ for the first moment. The leading-tail prediction
therefore reproduces the measured shell contributions to about $1\%$.
This agreement does not provide a rigorous bound on the remainder.

At $k=\pi/8$, the six-spinon spectrum contains 87 frequencies
in the main band and 19 in the high-energy tail. Relative sampling
errors in the main band range from $0.7\%$ to $2.8\%$.
Tail points with poorer statistical convergence are marked by
open symbols in Fig.~\ref{fig:spectra}.

Refining the main-band grid from 79 to 87 frequencies changes
the zeroth and first moments of the sampled spectrum by about
$1\%$. Tests over individual frequency intervals indicate that
further refinement is needed to establish the integration
accuracy. The companion article gives the frequency-grid
comparisons and sampling uncertainties for all eight momenta
\cite{companion}.

The upper-edge calculations in Fig.~\ref{fig:edges} integrate
over compact energy shells and require no rapidity cutoff,
providing an independent test of the analytical edge asymptotes.

\textit{Momentum-space TDVP.---}
We apply $\sum_j e^{\ii kj}S_j^z$ to the MPS ground state and express
the result as a left-gauge-fixed excitation ansatz
\cite{haegeman2012,haegeman2013} with a growing MPS window.
TDVP evolves this window \cite{haegeman2016,vandamme2022} using the
effective Hamiltonian of Ref.~\cite{osborne2025} and adapted environment
expansion \cite{mcculloch2024}. The ground-state background and leftmost
null-space tensor remain fixed. The window grows to the right when
the Frobenius overlap of its last tensor with the ground-state
tensor departs sufficiently from its background value.
For the initial bond-dimension-200 ground state, the correlation
length is $116.18$ and
$m_0^{\rm MPS}(\pi/2)=0.17008151312442$.
The first moment extracted from the initial time derivative agrees with
its exact value to about six parts per million.

The four-spinon time transform uses the dense frequency grid and
six spinons use direct full-band samples at $R=1040$. An independent
transform of the refined frequency grid gives a maximum phase discrepancy
of $0.115^\circ$ from TDVP over $0\le Jt\le15.25$.
For $C=x+\ii y$, sampling covariance
is propagated using gradients $(x,y)/|C|$ for amplitude and
$(-y,x)/|C|^2$ for phase, evaluated at the full partial sum.
These first-order bands omit TDVP and Bethe systematic errors.

% Let unused last-column space collect below the text, rather than in
% the display skips when REVTeX balances the final page.
\par\vfill\null
\label{em:end}

\end{document}